\documentclass[11pt,a4paper,onecolumn,reqno]{amsart}
\usepackage[a4paper, margin=2.3cm, bmargin=3cm]{geometry}

\usepackage{graphicx,stfloats} \usepackage{color}
\usepackage{amsmath}
\usepackage{amsaddr}
\usepackage{bm, soul}
\usepackage{mathtools}
\usepackage[colorinlistoftodos,prependcaption,textsize=tiny]{todonotes}
\usepackage{braket}
\usepackage{hyperref}
\usepackage[square,comma,sort&compress, numbers]{natbib}
\usepackage{enumerate}
\usepackage[version=4]{mhchem}

\usepackage{placeins}

\renewcommand{\st}[1]{}

\usepackage[square,numbers,sort&compress,comma]{natbib}

\usepackage[utf8]{inputenc}

\usepackage{amsmath}
\usepackage{amssymb}
\usepackage{caption}
\usepackage{graphicx}
\usepackage{latexsym}
\usepackage{times}

\usepackage{graphicx,stfloats}
\usepackage{color}

\usepackage{subcaption}
\usepackage{amssymb}
\usepackage{amsthm}
\usepackage{amsmath}
\usepackage{siunitx}

\usepackage{booktabs}
\newcolumntype{K}[1]{>{\centering\arraybackslash}p{#1}}
\newcolumntype{R}[1]{>{\raggedright\arraybackslash}p{#1}}

\definecolor{orange}{rgb}{1,.65,0}
\definecolor{lightgray}{gray}{0.9}

\usepackage{siunitx}
 
\usepackage{multirow}

\usepackage{etoolbox}
\patchcmd{\pprintMaketitle}
  {\hrule\vskip12pt}
 {\hrule\vskip12pt\ifvoid\extrainfobox\else\unvbox\extrainfobox\par\vskip12pt\fi}
 {}{}

\newsavebox\extrainfobox

\usepackage{caption}
\renewcommand{\figurename}{Fig.}

\title{Flamelet modeling of thermo-diffusively unstable hydrogen-air flames}

\author[stfs]{Hannes Böttler$^{1,*}$, Haris Lulic$^{1}$, Matthias Steinhausen$^{1}$, Xu Wen$^{1}$, Christian Hasse$^{1}$, Arne Scholtissek$^{1}$}
\email{boettler@stfs.tu-darmstadt.de} 
\address[]{$^1$Technical University of Darmstadt, Department of Mechanical Engineering, Simulation of reactive
Thermo-Fluid Systems, Otto-Berndt-Stra{\ss}e 2, 64287 Darmstadt, Germany
}

\begin{document}
\pagestyle{plain}

\maketitle

\begin{abstract}
In order to reduce \ce{CO2} emissions, hydrogen combustion has become increasingly relevant for technical applications.
In this context, lean \ce{H2}-air flames show promising features but, among other characteristics, they tend to exhibit thermo-diffusive instabilities.
The formation of cellular structures associated with these instabilities leads to an increased flame surface area which further promotes the flame propagation speed, an important reference quantity for design, control, and safe operation of technical combustors. While many studies have addressed the physical phenomena of intrinsic flame instabilities in the past, there is also a demand to predict such flame characteristics with reduced-order models to allow computationally efficient simulations.
In this work, a \ce{H2}-air spherical expanding flame, which exhibits thermo-diffusive instabilities, is studied with flamelet-based modeling approaches both in \textit{a-priori} and \textit{a-posteriori} manner.
A recently proposed Flamelet/Progress Variable (FPV) model, with a manifold based on unstretched planar flames, and a novel FPV approach, which takes into account a large curvature variation in the tabulated manifold, are compared to detailed chemistry (DC) calculations. 
Both flamelet approaches account for differential diffusion utilizing a coupling strategy which is based on the transport of major species instead of transporting the manifold control variables. 
First, both FPV approaches are assessed in terms of an \textit{a-priori} test with the DC reference dataset. Thereafter, the \mbox{\textit{a-posteriori}} assessment contains two parts: a linear stability analysis of perturbed planar flames and the simulation of the spherical expanding flame.
Both FPV models are systematically analyzed considering global and local flame properties in comparison to the DC reference data. 
It is shown that the new FPV model, incorporating large curvature variations in the manifold, leads to improved predictions for the microstructure of the corrugated flame front and the formation of cellular structures,
while global flame properties are reasonably well reproduced by both models.
\end{abstract}


\section{Introduction} \addvspace{10pt}

Due to the necessity to decarbonize modern economies, interest in hydrogen-fueled heat and power sources has increased, many of which will operate with lean premixed \ce{H2}-air flames. Despite their many advantages, lean \ce{H2} flames can be subject to intrinsic instabilities. To allow for safe and stable operation, the mechanisms and dynamics of the instabilities need to be understood and incorporated into appropriate modeling approaches which are used to design technical combustors. 

The most prominent instability mechanisms in premixed \ce{H2} flames are the thermo-diffusive (TD)  and the Darrieus-Landau (DL) instabilities. TD instabilities are caused by a disproportion of thermal and mass diffusion and can have either a stabilizing or destabilizing effect. TD-induced instabilities are only found for mixtures with Lewis numbers smaller than unity. 
DL instabilities on the other hand originate from thermal expansion and the associated density change across the flame front and always have a destabilizing effect for all flames. 
Recent studies have investigated intrinsic instabilities by means of experiments~\cite{kwon_2002,hall_2015,bauwens_2017,fernandez_galisteo_2018}, asymptotic theory~\cite{creta_2020} and direct numerical simulations (DNSs)~\cite{altantzis_2012, altantzis_2015, frouzakis_2015, berger_2019, attili_2021, howarth_2022}. All of these studies have shown that the overall flame propagation speed is enhanced by flame surface corrugations originating from instabilities. 
Furthermore, thermo-diffusively unstable flames tend to exhibit characteristic cell sizes. In order to investigate such instabilities, the linear stability analysis has been utilized by determining the so-called dispersion relation from either asymptotic theory~\cite{creta_2020} or numerical simulations~\cite{altantzis_2012, altantzis_2015, berger_2019}.
A comprehensive review and a theoretical introduction to intrinsic flame instabilities can be found in~\cite{howarth_2022}.

So far, numerical studies of intrinsic flame instabilities have usually been based on DNS. It has been shown that reliable simulations addressing flame instabilities require considerable computational resources since the formation and evolution of intrinsic instabilities can be sensitive to domain size, grid resolution, and the numerical methods employed~\cite{yu_2017,berger_2019,howarth_2022}. Therefore, such investigations are restricted to academic configurations, which are also well-suited for model development. 
Promising modeling approaches, taking into account detailed kinetics and their interaction with transport, include flamelet-based models such as the Flamelet/Progress Variable (FPV) approach~\cite{pierce_2004}, Flamelet Generated Manifolds (FGM)~\cite{oijen_2016}, or the Flame Prolongation for ILDM (FPI) model~\cite{gicquel_2000}. 
While manifolds for non-premixed combustion are usually generated from stretched non-premixed flamelets~\cite{peters_1984}, manifolds for premixed combustion can recover moderate stretch effects even when being generated from unstretched premixed flamelets~\cite{oijen_2016, schlup_2019, mukundakumar_2021, boettler_2021b}. Nevertheless, attempts have also been made to generate manifolds from stretched premixed flamelets~\cite{oijen_2010,knudsen_2013,oijen_2016}.
Many advancements were aiming for improved predictions of the local mixture composition which has been realized via mixture fraction definitions based on elemental mass fractions, such as the Bilger mixture fraction, and efforts have been made to model its diffusivity~\cite{bilger_1990, oijen_2010, oijen_2016}.
Recently, different model extensions have been developed to improve the predictions of flamelet-based models for hydrogen combustion, where differential diffusion effects need to be captured accurately~\cite{schlup_2019, mukundakumar_2021, boettler_2021b}. 
For TD unstable flames, which exhibit large curvature variations due to flame front corrugations, the suitability of these manifolds needs to be further investigated.
A novel flamelet tabulation approach, which includes strain and curvature effects, has shown reasonable agreement with the detailed chemistry (DC) result of a lean hydrogen spherical expanding flame (SEF)~\cite{wen_2021a}. The manifold was evaluated by extracting the control parameters from the DC simulation and comparing the tabulated thermochemical state to the reference (\textit{a-priori} analysis).
To the authors' knowledge, fully coupled simulations (\textit{a-posteriori} analysis) of intrinsically unstable \ce{H2}-air flames using flamelet manifolds which incorporate large curvature variations have not been presented in the literature, yet. This gap is addressed in this work. 

The objective of this study is twofold: (1)~a novel flamelet tabulation approach is presented which takes into account differential diffusion and large curvature variations (positive and negative curvature range). It is studied together with a previously developed manifold constructed from unstretched premixed flames~\cite{boettler_2021b}. The predictive capabilities of both manifolds are first assessed by comparison to DC reference results of the unstable lean \ce{H2}-air SEF also studied by Wen et al.~\cite{wen_2021a} (\textit{a-priori} analysis). (2)~The new manifold is then coupled to a CFD solver in a modified FPV approach for premixed combustion and utilized to simulate planar flames and the lean \ce{H2}-air SEF (\textit{a-posteriori} analysis). The FPV modeling approach from our previous work~\cite{boettler_2021b} is employed in a similar manner. 
Both FPV models are examined \textit{a-posteriori} by means of a linear stability analysis (planar flames) and the global characteristics of the SEF, as well as its cellular microstructure.  These flames represent challenging cases for flamelet-based modeling approaches due to their highly unsteady nature and distinct flame front corrugations (large curvature variations).

The paper is structured as follows: the new tabulation approach and the numerical model are introduced first, followed by the \textit{a-priori} analysis. Thereafter, fully coupled simulations are performed for the planar flames (linear stability analysis) and the SEF. The paper ends with a conclusion.

\section{Numerical method} \addvspace{10pt}
Two flamelet-based modeling approaches are investigated: (1)~a recently proposed FPV approach using a three-dimensional manifold ($\psi(Z_\mathrm{Bilger}, h, Y_c)$) which was created from unstretched flame solutions with varying enthalpy levels $h$~(referred to as \mbox{FPV-$h$})~\cite{boettler_2021b} and (2)~a novel three-dimensional tabulation approach taking into account negative and positive curvature $\kappa_c$~(referred to as FPV-$\kappa_c$). First, the novel tabulation approach is presented, followed by a description of the numerical setup of the unstable lean \ce{H2}-air SEF. Note that, in all calculations, a mixture-averaged diffusion model~\cite{curtiss_1949} without thermal diffusion and the detailed reaction mechanism by Varga et al.~\cite{varga_2015} are used in accordance with the DC SEF model~\cite{wen_2021a}.

\subsection{Manifold generation (FPV-$\kappa_c$)} \addvspace{10pt}
Flamelet-based manifolds are usually generated by performing parameter variations in simple laminar canonical flame configurations, such as freely propagating flames or counterflow flames. However, these laminar canonical flames are inherently limited regarding the attainable parameter space of stretch (strain, curvature)~\cite{boettler_2021}. 
To overcome this limitation, a composition space model (CSM)~\cite{scholtissek_2019b} is used instead of a specific laminar canonical configuration. As shown in our previous work, the CSM recovers the characteristics of various laminar premixed flames and can further take into account arbitrary combinations of strain $K_s=\nabla_t\cdot\mathbf{u}_t - \left(\mathbf{u}\cdot\mathbf{n}_c\right)\kappa_c$, and curvature, $\kappa_c=-\nabla\cdot\mathbf{n}_c$ where $\nabla_t\cdot\mathbf{u}_t$ represents flame-tangential straining by the
flow and $\mathbf{n}_c$ is the flame-normal unit
vector ~\cite{boettler_2021}. In the model equations for the temperature, species mass fractions, and progress variable gradient are expressed and solved with respect to the progress variable $Y_c$, which spans the composition space. With the gradient equation, the model is self-contained, requiring only strain and curvature as external parameters~\cite{scholtissek_2019b}. 

\begin{figure}[ht]
\centering
\vspace{-7pt}
\includegraphics[scale=0.95]{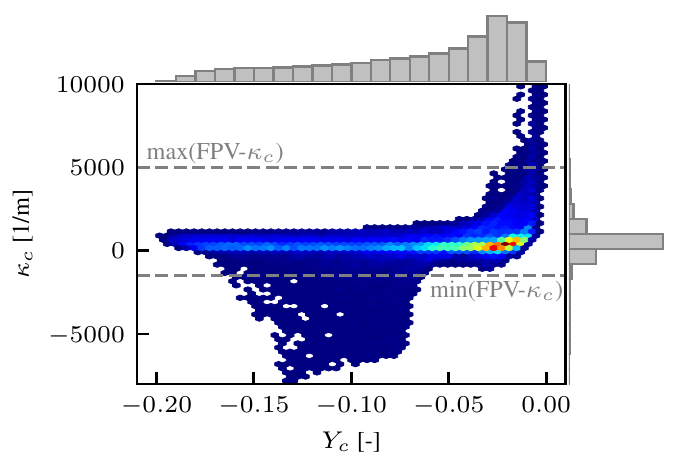}
\caption{Joint probability density function (PDF) of progress variable and curvature evaluated from the DC data on the SEF (presented in Sec.~2.2) with qualitative PDFs shown as bars at the respective edges. Dashed lines indicate the curvature range captured in the FPV-$\kappa_c$ manifold.}
\label{fig:jpdf_yc_curvature}
\end{figure}

In order to generate the manifold, a parameter variation is carried out for a range of equivalence ratios ($0.275 \leq \phi \leq 0.6$) and curvatures ($\SI{-1500}{ 1 \per m} \leq \kappa_c \leq \SI{5000}{ 1 \per m}$). The strain rate is fixed at $K_s = \SI{500}{ 1 \per s}$ to account for the positive strain of the SEF during its outward propagation (numerical setup presented in Sec.~2.2). The progress variable is defined as $Y_c=Y_{\ce{H2O}}-Y_{\ce{H2}}-Y_{\ce{O2}}$. Figure~\ref{fig:jpdf_yc_curvature} shows a joint probability density function (PDF) of the $\kappa_c$ -- $Y_c$ scatter which is evaluated from the DC data of the SEF. 
For visual inspection, qualitative PDFs of both quantities are depicted at the respective axes. The curvature range of the FPV-$\kappa_c$ manifold is indicated by dashed lines. It is noted that the curvature distribution of the DC reference is well captured, while few regions can be identified where the tabulated range is exceeded. However, no burning solutions could be generated with the CSM below $\kappa_c=\SI{-1500}{1 \per \meter}$ for this type of lean \ce{H2}-air flame.

\begin{figure}[t!]
\centering
\includegraphics[scale=0.95]{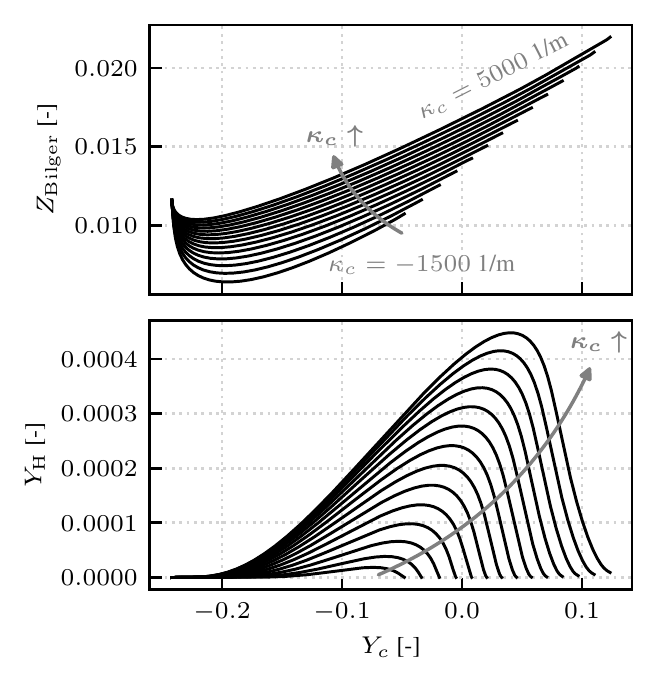}
\caption{Subset of the CSM solutions used for manifold generation with $\phi=0.4$ ($Z_\mathrm{Bilger}=0.0116$) and varying curvature ($\SI{-1500}{ 1 \per m} \leq \kappa_c \leq \SI{5000}{ 1 \per m}$).}
\vspace{-4pt}
\label{fig:tabulated_flamelets}
\end{figure}

Highlighting the effect of curvature variation, CSM solutions with $\phi=0.4$ ($Z_\mathrm{Bilger}=0.0116$) are \mbox{illustrated} in \figurename\ref{fig:tabulated_flamelets}. 
The profiles of the Bilger mixture fraction $Z_\mathrm{Bilger}$~\cite{bilger_1990} and \ce{H} radical mass fraction are shown as a function of the progress variable.
Curvature variations are found to lead to different local mixture compositions and maximum values of $Y_c$ due to pronounced differential diffusion effects. Further, the \ce{H} radical profile shows a monotonic increase with curvature, which is in line with previous works that have used this quantity to include stretch effects in flamelet-based approaches~\cite{oijen_2010,knudsen_2013, wen_2021a}.

The flamelet manifold is obtained first by mapping the set of flame calculations ($\psi(\phi, Y_c, \kappa_c)$) to $Z_\mathrm{Bilger}$, which is defined by a coupling function between the fuel (subscript~1) and oxidizer (subscript~0)~\cite{bilger_1990}:
\begin{equation}
    Z_\mathrm{Bilger} = \frac{\beta - \beta_0}{\beta_1 - \beta_0}
\end{equation}
where the coupling functions $\beta$ depend on the elemental mass fractions $Z_l$,
\begin{equation}
    \beta = \sum_{l=1}^{N_e} \gamma_l \sum_{k=1}^{N_s} \frac{a_{l,k} M_l Y_k}{M_k}
\end{equation}
with the weighting factor $\gamma_l$ of element $l$, the molecular weight $M_l$ ($M_k$) of element $l$ (species $k$), and $a_{l,k}$ representing the number of elements $l$ in species $k$. Note that $\gamma_l$ are chosen in agreement with Bilger et al.~\cite{bilger_1990}.

Second, the progress variable $Y_c$ is mapped to a normalized progress variable:
\begin{equation}
c = \frac{Y_c - Y_{c,\mathrm{min}}(Z_\mathrm{Bilger})}{Y_{c,\mathrm{max}}(Z_\mathrm{Bilger})- Y_{c,\mathrm{min}}(Z_\mathrm{Bilger})}
\end{equation}
Finally, the curvature dimension is mapped to a second normalized progress variable based on the \ce{H} radical:
\begin{equation}
c_2 = \frac{Y_{\ce{H}} - Y_{\ce{H},\mathrm{min}}(Z_\mathrm{Bilger}, c)}{Y_{\ce{H},\mathrm{max}}(Z_\mathrm{Bilger}, c)- Y_{\ce{H},\mathrm{min}}(Z_\mathrm{Bilger}, c)}
\end{equation}
The tabulated manifold ($\psi(Z_\mathrm{Bilger}, c,c_2)$) is coupled to the CFD solver following the approach presented in our previous work~\cite{boettler_2021b} by transporting the major species (\ce{H2O}, \ce{H2}, \ce{O2}) and approximating the control variables of the tabulated manifold based on these species. 
For the FPV-$h$ model, the enthalpy is transported and used as a control variable for the manifold instead of the \ce{H} radical. The manifold is generated with the CSM performing an enthalpy variation for unstretched flames and mapping the thermochemical state to $\psi(Z_\mathrm{Bilger}, h, Y_c)$ (further details are provided in~\cite{boettler_2021b}). Both coupling approaches avoid the use of additional modeling assumptions which would otherwise be required to approximate the transport \mbox{properties} of composed quantities such as the mixture fraction or the progress variable.
Finally, it is noted that these coupling approaches are flexible in number of species being transported as long as a consistent mixture fraction definition is used~\cite{boettler_2021b}.

\begin{figure*}[hb]
\centering
\vspace{-5pt}
\includegraphics[scale=0.945]{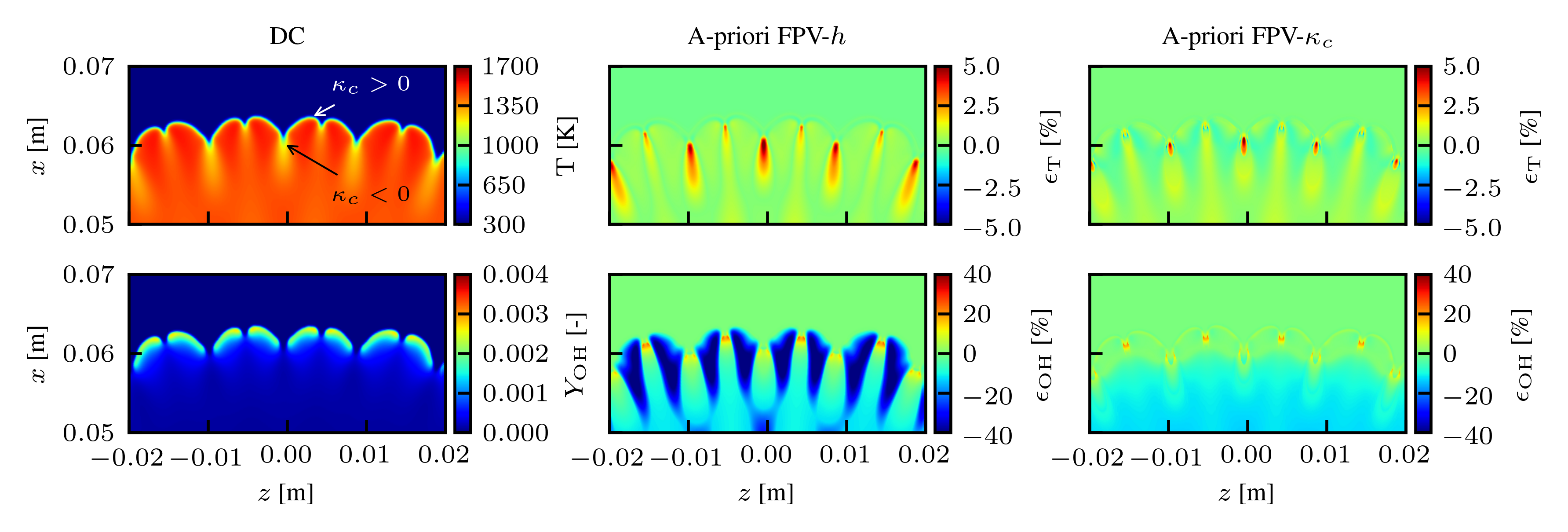}
\caption{\textit{A-priori} comparison of FPV approaches based on DC SEF data. The absolute DC values are shown for the temperature $T$ and \ce{OH} mass fraction $Y_\mathrm{OH}$ (left), while deviations in the predictions of the FPV-$h$ (middle) and the FPV-$\kappa_c$ model (right) are shown relative to these values (denoted as $\epsilon_i = (i_\mathrm{DC} - i_\mathrm{FPV}) / \mathrm{max}(i_\mathrm{DC})$).}
\label{fig:apriori}
\vspace{-6pt}
\end{figure*}

\subsection{Numerical setup} \addvspace{7pt}

A lean premixed spherical expanding \ce{H2}-air flame is used to investigate the performance of the FPV models in predicting cellular flame structures which evolve at larger flame radii. For this purpose, the DC reference data, which was already investigated in an \textit{a-priori} analysis by Wen et al.~\cite{wen_2021a}, is used for comparison. This dataset was compared against asymptotic theory~\cite{wen_2021b} showing good agreement with the reference data. It is further noted that the DC framework was validated against experimental data where only a slight under prediction of the consumption speed was observed~\cite{fzhang_2017}.
The computational domain is a two-dimensional, axisymmetric wedge with a radius of $R=\SI{8}{cm}$. In the domain center, a hot spot with a radius of \SI{1}{cm} is initialized. Its conditions correspond to the equilibrium state of the fresh gases, a \ce{H2}-air mixture with an equivalence ratio $\phi=0.4$ and an unburnt temperature $T_\mathrm{u}=\SI{300}{K}$. Outside of the hot spot, the domain is initialized with the unburned mixture at atmospheric pressure. 
Local grid refinement ensures that the flame is resolved with at least $20$ grid points. It has been shown that computations of unstable \ce{H2}-air flames can depend significantly on the grid resolution, initialization, and numerical setup~\cite{yu_2017,howarth_2022}. Therefore, the FPV calculations are based on the same numerical setup as the DC calculation. Additionally, to ensure a consistent initialization for the FPV calculations, an early time step of the DC simulation is used, where the flame front is fully developed and flame front corrugations are still below \SI{4}{\percent} of the laminar flame thickness $l_\mathrm{f}$.

\section{Results and Discussion}

\subsection{A-priori: Spherical expanding flame}\addvspace{10pt}
An \textit{a-priori} analysis is carried out to assess the general capability of the FPV models to capture the microstructure of the unstable \ce{H2}-air flame. Figure~\ref{fig:apriori} shows a snapshot of the temperature $T$ and the \ce{OH} mass fraction $Y_\mathrm{OH}$ fields of the DC simulation (left). Further, the relative deviations of both FPV model predictions from the DC reference are shown. 
For both models, the highest deviations in the temperature can be found around areas of negative flame curvature (concave toward the unburned mixture), whereas smaller deviations can be found in positively curved segments (convex toward the unburned mixture). Similar, but more significant deviations are registered for the \ce{OH} mass fraction. While the FPV-$h$ model shows maximum deviations of up to \SI{5}{\%} for $T$ and \SI{40}{\%} for $Y_{\mathrm{OH}}$, the prediction of the FPV-$\kappa_c$ approach exhibits smaller deviations for both quantities. The deviations at negatively curved regions are also restricted to a smaller area of the flame in comparison to the FPV-$h$ manifold. 
This indicates that taking curvature effects into consideration in the \mbox{FPV-$\kappa_c$} approach improves the agreement for the unstable SEF, especially for fine-scale quantities such as the \ce{OH} radical, which is particularly challenging for flamelet-based models due to the strong differential diffusion effects and broad curvature distribution. An \textit{a-priori} assessment of flame-tangential diffusion effects, which supports the previous line of argument, can be found in the supplementary material.

\subsection{A-posteriori: Linear stability analysis}\addvspace{10pt}

A comprehensive assessment of the performance of flamelet-based approaches requires \mbox{\textit{a-posteriori}} analyses of fully coupled simulations, which are discussed in the following. The \mbox{\textit{a-posteriori}} assessment is divided into two parts: (1) the linear stability analysis of perturbed planar flames and (2) the spherical expanding flame.

In order to examine the cell formation predicted by the different FPV models, the linear stability analysis (LSA) has proven to be a viable method for studying flame instabilities~\cite{frouzakis_2015, berger_2019}.  
An LSA is performed simulating fully developed two-dimensional planar flames in a box, subject to a weak initial perturbation with an initial amplitude $A_0=0.04 \,l_\mathrm{f}$ and the wavelength $\lambda$. The two-dimensional domain has inflow and outflow conditions in the streamwise direction and periodic boundaries in the lateral direction.  
\begin{figure}[ht]
\centering
\vspace{-3pt}
\includegraphics[scale=1]{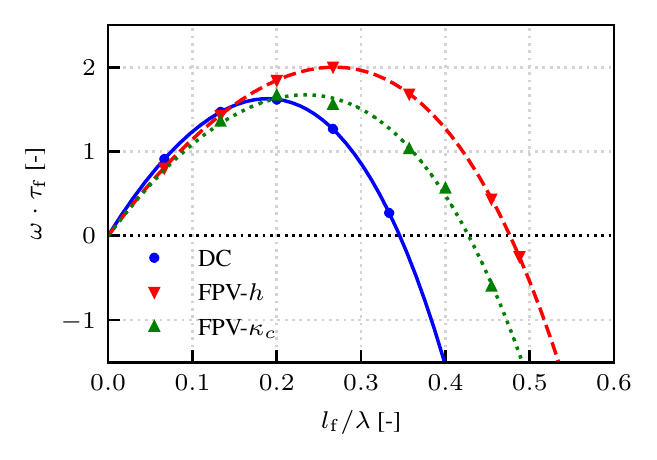}
\caption{Comparison of the non-dimensional dispersion relations obtained with detailed chemistry (DC) and the two FPV models, respectively.}
\label{fig:dispersion_relation}
\vspace{-10pt}
\end{figure}
The growth rate $\omega$ is calculated for each wavelength as $\omega=\mathrm{d}\left(\ln{A(t)}/{A_{0}}\right)/\mathrm{d}t$. The domain size in the direction of flame propagation is large enough to permit unconstrained flame propagation, while its lateral dimension is varied to adjust $\lambda$. The dispersion relation is obtained as $\omega=\omega(k)$, where $\omega$ is the growth rate of the initial perturbation defined by the wavelength $\lambda$ and the wavenumber $k=2\pi/\lambda$.

The dispersion relation reveals the range of unstable (and stable) wavelengths, the critical (neutral, $\omega=0$) wavelength $\lambda_c$, and the most unstable wavelength $\lambda(\omega_\mathrm{max})$ at which the growth rate reaches a maximum. The dispersion relation includes both hydrodynamic and thermo-diffusive effects similar to the unstable SEF, the main subject of this study.

The dispersion relations obtained from detailed chemistry (DC) and both FPV methods are shown in dimensionless form in \figurename\ref{fig:dispersion_relation}. For normalization, the flame thickness $l_\mathrm{f} = (T_\mathrm{b}-T_\mathrm{u}) / \mathrm{max}(\mathrm{d}T/\mathrm{d}x)$ and the flame time $\tau_\mathrm{f}=l_\mathrm{f}/s_\mathrm{L}$ of the unperturbed freely propagating flame are used, where $s_\mathrm{L}$ represents the laminar burning velocity. 
A wider range of unstable wavelengths is recorded for both FPV models in comparison to the DC solution, with a smaller deviation for the FPV-$\kappa_c$ model. The maximum growth rate is reproduced well by the FPV-$\kappa_c$ approach and the peak position $\lambda(\omega_\mathrm{max})$ is slightly shifted towards lower wavelengths. For the FPV-$h$ model, the peak value of the growth rate is larger and its position is shifted to smaller wavelengths. The overall shape of the curve is captured well by both models and a good agreement is found for larger wavelengths.
The shift in the peak position and range of wavelengths for the FPV models provides insights into the length scales of instabilities that will prevail and grow in premixed flame fronts. 
\begin{figure*}[ht]
\centering
\vspace{-10pt}
\includegraphics[scale=0.99]{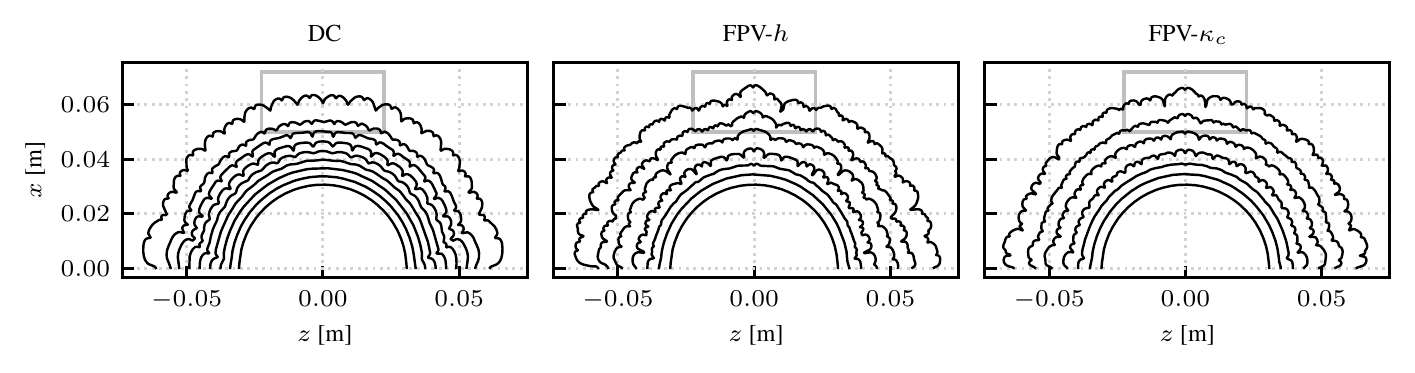}
\caption{Comparison of flame fronts obtained by different simulation approaches and various time steps. The flame front is defined as the progress variable iso-line $Y_c=-0.11$. The gray boxes visualize the magnified region which will be investigated in detail in \figurename\ref{fig:compare_flamestructure}.}
\vspace{-18pt}
\label{fig:compare_flamefronts}
\end{figure*}
It has been found in several studies of planar~\cite{berger_2019,altantzis_2012} 
and circular expanding flames~\cite{altantzis_2015} that the average cell size in these unstable flames is related to the fastest growing modes close to $\lambda(\omega_\mathrm{max})$. 
It can therefore be concluded from the dispersion relations in \figurename\ref{fig:dispersion_relation} that smaller cells are enhanced with the FPV-$\kappa_c$ model and even smaller ones with the \mbox{FPV-$h$} model as compared to the detailed chemistry result. These findings are examined further for the SEF configuration in the following.

\subsection{A-posteriori: Spherical expanding flame}\addvspace{10pt}

Next, the performance of the FPV models is assessed by analyzing the growth of instabilities in the SEF configuration.
The flame front evolution of the SEF for all three simulation approaches is depicted in \figurename\ref{fig:compare_flamefronts}. The flame front is defined based on the progress variable isoline $Y_c=-0.11$ in accordance with~\cite{wen_2021a}. 

All calculations are performed until a flame radius of approximately \SI{6}{\centi \meter} is reached. Despite having similar overall characteristics, all three approaches lead to slightly different flame shapes. In the DC simulation, cellular structures of approximately uniform size evolve.
Similarly sized cells are also found in the flame front obtained from the \mbox{FPV-$h$} calculation; however, smaller secondary cellular structures are also seen to form.
This leads to a more corrugated flame front at larger flame radii. 
The visual inspection of the results obtained with the \mbox{FPV-$\kappa_c$} model indicates a similar formation of secondary cells as found for the \mbox{FPV-$h$} model. Nevertheless, the flame front predicted by the \mbox{FPV-$\kappa_c$} model is less strongly affected by small cells compared to the \mbox{FPV-$h$} approach. Overall, this observation is in agreement with the dispersion relations in \figurename\ref{fig:dispersion_relation} since the FPV models generally predict smaller critical wavelengths, with the \mbox{FPV-$\kappa_c$} model showing better agreement with the DC reference.

Further, the unstable wavelengths obtained by the dispersion relation are quantitatively compared to the size of cellular structures found in the SEFs by estimating cell size distributions. The cell size is defined as $\lambda_\mathrm{cell} = 2 \, l_\mathrm{arc} / \pi$ where the arc length $l_\mathrm{arc}$ is defined as the distance between two curvature peaks along the flame front~\cite{berger_2019}. 
\begin{figure}[ht]
\vspace{-6pt}
\includegraphics[scale=0.98]{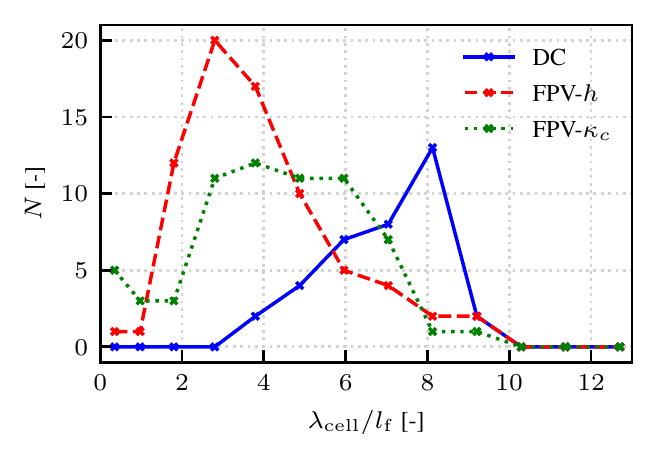}
\caption{Comparison of the non-dimensional cell size distributions predicted by different modeling approaches.}
\label{fig:compare_cell_size}
\vspace{-2pt}
\end{figure}
Figure~\ref{fig:compare_cell_size} shows the cell size distributions of the different SEF simulations. Note that the cell size is normalized by the laminar flame thickness $l_\mathrm{f}$ and $N$ represents the number of cells found in each bin.
A maximum is visible in the distribution of all models which can be related to the most unstable wavelength $\lambda(\omega_\mathrm{max})$~\cite{berger_2019}. The maximum occurs at smaller cell sizes for \mbox{FPV-$\kappa_c$} and \mbox{FPV-$h$} in comparison to the DC reference, which is also consistent with the shift in the maximum growth rate in the corresponding dispersion relations~(see \figurename\ref{fig:dispersion_relation}). Similarly, the critical wavelength is shifted to smaller values for the FPV models, which is reflected by the occurrence of smaller cells in the distributions. 
Berger et al.~\cite{berger_2019} found that the most likely wavelength of their investigated lean \ce{H2}-air flames occurs around $6 \, l_\mathrm{f}$. With $7-8 \, l_\mathrm{f}$, this value is larger for the SEF computed with the DC simulation and differences are attributed to the overall curved flame configuration and the limited sample size as compared to~\cite{berger_2019}.

\begin{figure}[ht]
\centering
\vspace{-2pt}
\includegraphics[scale=0.98]{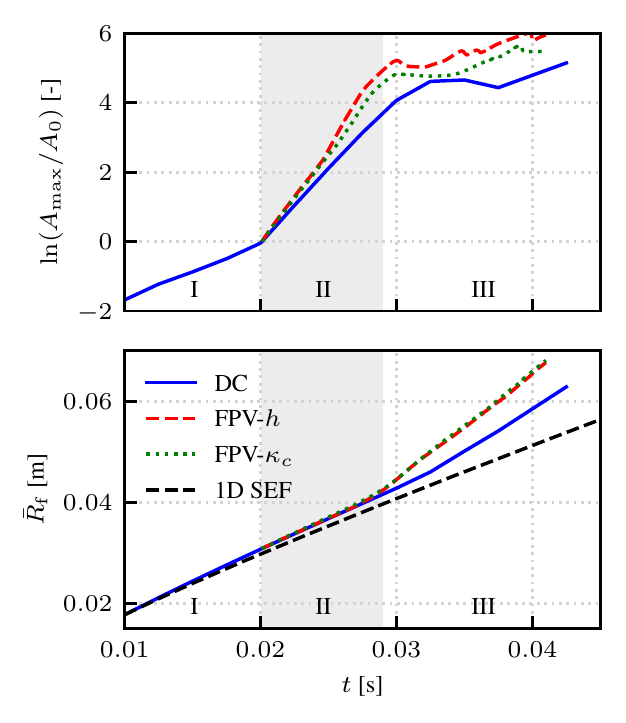}
\caption{Temporal evolution of the maximum amplitude of perturbation $A_\mathrm{max} = R_\mathrm{f,max} - R_\mathrm{f,min}$ obtained from SEF simulations with respective models normalized by the initial perturbation $A_0$ (top).  Further, the temporal evolution of the mean flame radius $\bar{R}_\mathrm{f}$ is depicted for the respective calculations (bottom). Additionally, the regimes of unperturbed propagation (I), linear growth (II), and non-linear growth (III) are highlighted and the DC result of a one-dimensional SEF simulation is shown for reference as no instabilities can evolve due to the spatial confinement (black line).}
\vspace{-5pt}
\label{fig:compare_amplitude+propagation}
\end{figure}

The differently sized cells can also be explained by
analyzing the temporal evolution of the maximum perturbation $A_{\mathrm{max}}$ 
along the mean flame radius $\bar{R}_\mathrm{f}$
depicted in \figurename\ref{fig:compare_amplitude+propagation} (top).
Three different regimes can be identified for the DC reference. Initially, an unperturbed flame propagation, with negligible perturbations of the flame front, is identified (regime~I), followed by a linear regime (regime~II), which can be related to the linear stability analysis of the planar flames, and a non-linear regime (regime~III), which is characterized by the interaction and chaotic superposition of different cellular structures. 
Note that the FPV calculations are initialized from the DC calculation at the beginning of regime~II ($t=\SI{0.02}{ms}$). Here, the initial amplitude of perturbation $A_0$ is smaller than $\SI{4}{\percent} \, l_\mathrm{f}$, which is similar to the initial corrugations prescribed in the LSA.
Thereby, consistency between the DC calculation and the FPV models is ensured since smaller perturbations might be sensitive to the different numerical models. 

In regime~II, all approaches show a linear trend, while the FPV models show a faster increase (larger growth) compared to the DC reference. This is in agreement with the overall picture obtained for the flame fronts predicted by the FPV calculations, since an increased growth of perturbation leads to a faster development of cellular structures. 
The final regime~III is characterized by moderate growth in the amplitude of the perturbation since it is characterized by the interaction of different cells.
This confirms the initial qualitative assessment of the flame front evolution~(see \figurename\ref{fig:compare_flamefronts}). 

\begin{figure*}[hb]
\centering
\vspace{-6pt}
\includegraphics[scale=0.99]{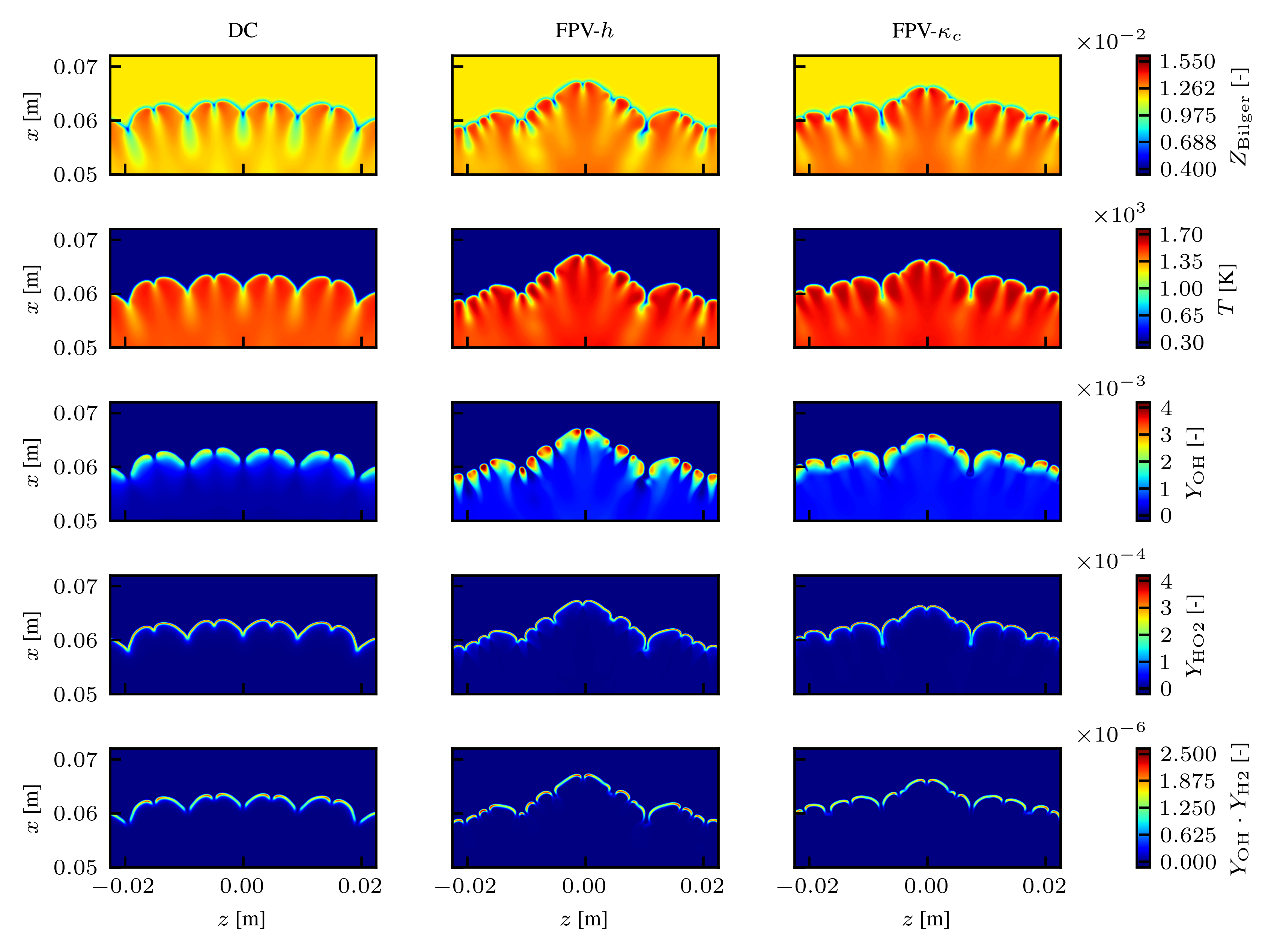}
\caption{Flame structure comparison for the DC reference, the FPV-$h$, and the FPV-$\kappa_c$ model at a mean flame radius of $\bar{R}_\mathrm{f}\approx\SI{6}{\centi\meter}$. The sections of the flame front are color-coded by $Z_\mathrm{Bilger}$, $T$, $Y_\mathrm{OH}$, $Y_\mathrm{HO2}$, and the product $Y_\mathrm{OH} \cdot Y_\mathrm{H2}$ with color bars given for each row.}
\vspace{-10pt}
\label{fig:compare_flamestructure}
\end{figure*}

The flame propagation speed is enhanced by the wrinkling due to intrinsic instabilities \cite{altantzis_2015, berger_2019, creta_2020}. This is re-confirmed in \figurename\ref{fig:compare_amplitude+propagation} (bottom), where the temporal evolution of the mean flame radius $\bar{R}_\mathrm{f}$ is shown for the different models. For reference, a one-dimensional SEF is shown, computed with an in-house flame solver~\cite{zschutschke_2017}. Due to the confinement to a single dimension, no instabilities can evolve and the flame propagates in a quasi-steady manner.
While the two-dimensional simulations predict a similar flame evolution as the one-dimensional model in the first two regimes, a steeper increase $\mathrm{d}\bar{R}_\mathrm{f}/\mathrm{d}t = s_\mathrm{f}$ (i.e.~a higher flame speed) is observed as soon as cells of different size interact (regime III). 
Due to the more pronounced formation of smaller cellular structures in the FPV calculations (thus, also a larger flame surface area), a faster flame propagation is predicted by these models in comparison to the DC simulation. 
Considering the complexity of this particular flame with differential diffusion, curvature effects, unsteadiness, and the evolution cellular structures, this agreement is quite remarkable. Flame speed deviations of \SI{15}{\percent} are observed for flames with the largest radius. 
A similar overprediction in flame propagation speed was observed in a previous study using a two-dimensional unstretched manifold ($\psi(Z,Y_c)$)~\cite{schlup_2019}.

So far, the results have shown that the model predictions of the instability evolution affects both global flame characteristics, such as the flame propagation speed, as well as local flame characteristics, such as the cell size distribution and visual appearance of the flame. As a next step, the results of the coupled simulations are analyzed with respect to the flame microstructure. Clear differences are found in the \textit{a-priori} analysis (see \figurename\ref{fig:apriori}) and a similar investigation is carried out for the \textit{a-posteriori} results. 
The thermo-chemical states obtained with the different models are compared in \figurename\ref{fig:compare_flamestructure}. 
Here a magnified picture of the flame front is depicted (gray boxes in \figurename\ref{fig:compare_flamefronts}) and the scalar fields of the Bilger mixture fraction $Z_\mathrm{Bilger}$, temperature $T$, and \ce{OH} mass fraction $Y_\mathrm{OH}$ are shown.
Richer (leaner) mixtures are found in regions of positive (negative) curvature, which is expected for lean \ce{H2}-air flames. While both FPV models show similar mixture stratification across the flame, they tend to predict richer mixtures on the burned side of the flame in comparison to the DC calculation. This also leads to a slight overprediction of the temperature on the burned side. On the other hand, areas of leaner mixtures due to negative curvature seem to be less prevalent in the FPV calculations compared to the DC simulation. However, this effect depends on the size of the cellular structures, since smaller cells lead to higher curvature variation which also amplifies differential diffusion effects.
Finally, the \ce{OH} mass fraction is shown, which scales with the local heat release, indicating varying local reaction intensities originating from curvature effects. From the DC snapshot, \ce{OH} is seen to exhibit the highest values in the positively curved flame front while no significant \ce{OH} mass fraction is found in negatively curved segments, highlighting its sensitivity to curvature. These characteristics are also found in the FPV simulations. 
The FPV-$h$ model generally predicts higher \ce{OH} mass fractions and shows a broader area with a significant \ce{OH} content in comparison to the DC reference. Moreover, the FPV-$\kappa_c$ result agrees better with the DC calculation. It only slightly overpredicts $Y_\mathrm{OH}$ in the positively curved flame segments and indicates the reaction zone thickness is similar to the DC reference. Further, the \ce{HO2} mass fraction is depicted which exhibits non-negligible values also in regions of negative curvature. Similar observations were reported by Hall et al.~\cite{hall_2016a}. This is consistently observed in all three modeling approaches, highlighting that weak reactions still occur in areas of negative curvature.
Finally, also the product of \ce{H2} and \ce{OH} mass fractions is shown since it was found to be a suitable marker for heat release in lean unstable \ce{H2}/air flames~\cite{marshall_2019}. High (low) heat release is found for areas with positive (negative) curvature for the three different modeling approaches, respectively. In general, all characteristics of the DC reference are reproduced by the FPV models, while the FPV-$h$ model shows a slight overprediction in positively curved regions.
Overall, these results are in agreement with the \textit{a-priori} analysis performed initially. While global flame characteristics do not deviate significantly between both FPV models, the local flame characteristics and the flame's microstructure predicted by the FPV-$\kappa_c$ model agrees better with the DC simulation. 

\section{Conclusion} \addvspace{10pt}

In this work, a lean \ce{H2}-air spherical expanding flame (SEF), which exhibits thermo-diffusive instabilities, is studied with flamelet-based modeling approaches both in \textit{a-priori} and \textit{a-posteriori} manner.
A recently proposed \mbox{FPV-$h$} model~\cite{boettler_2021b}, with a manifold based on unstretched planar flames, and a novel \mbox{FPV-$\kappa_c$} modeling approach, which takes into account a large curvature variation in the tabulated manifold, are compared to detailed chemistry (DC) calculations of the hydrogen flame. Furthermore, a linear stability analysis (LSA) is performed in order to systematically determine and compare the growth rates of premixed flame perturbations predicted by the three modeling approaches (\mbox{FPV-$h$}, \mbox{FPV-$\kappa_c$}, DC).

Overall, the comparison of both FPV approaches to the DC reference shows that incorporating curvature in the manifold (\mbox{FPV-$\kappa_c$}) leads to more accurate predictions for the local characteristics and the microstructure of the flame instabilities than the manifold based on unstretched flames (\mbox{FPV-$h$}). The \mbox{\textit{a-priori}} analysis reveals that the \mbox{FPV-$\kappa_c$} manifold leads to good predictions of the thermo-chemical state, especially in negatively curved flame segments, where the \mbox{FPV-$h$} model shows significantly larger deviations. The aspect is further confirmed from the \textit{a-posteriori} results for the cell size distributions and the flame structure comparison of the flame with developed cellular structures.
Additionally, as indicated by the dispersion relations obtained from the LSA, the coupled simulations show that the flames computed with both FPV approaches form smaller \mbox{cellular} structures, corresponding to smaller critical wavelengths, compared to the SEF-DC reference model. This increases the flame surface area and thereby also the flame propagation speed, an important global flame characteristic. While both FPV approaches recover the unstretched laminar burning velocity very accurately, the overall agreement of the SEF propagation speed between the FPV models and the DC simulation is still remarkable, given the challenging nature of the flame physics involving transient flame propagation, dynamic instability evolution, large curvature variation, and differential diffusion. Hence, the coupling strategy utilized in this work, which is based on the transport of major species rather than transporting the manifold control variables, shows a high potential for future applications.
Additionally, an extension of this study to increased pressure levels would be a beneficial contribution to the field.


\section*{Acknowledgments} \addvspace{10pt}
The research leading to these results has received funding from the European Union’s Horizon 2020 research and innovation program under the Center of Excellence in Combustion (CoEC) project, grant agreement No 952181 and from the German Research Foundation (DFG) - Project No. 411275182.
HL acknowledges funding by the Fritz and Margot Faudi-Foundation - Project No. 55200502.
Numerical simulations were conducted on the Lichtenberg II High Performance Computer of the Technical University of Darmstadt.
The authors thank T. Zirwes for providing the DC dataset of the SEF.

\bibliography{publication.bib}
\bibliographystyle{unsrtnat_mod}

\end{document}


\pagestyle{plain}

\maketitle

This supplementary material contains an assessment of tangential diffusion effects in the thermo-\-diff\-usively unstable \ce{H2}/air spherical expanding flame in an \textit{a-priori} manner. The diffusive fluxes of selected species are evaluated on the DC data together with corresponding predictions of the manifolds, respectively. Further, the different models are compared with respect to a coordinate measured along the corrugated flame front. In general, the FPV-$\kappa_c$ model gives a more accurate prediction of the diffusive fluxes in flame-normal and flame-tangential direction compared to the FPV-$h$ model. This is in agreement with the \textit{a-priori} analysis provided in Section~3.1 of the manuscript, since the FPV-$\kappa_c$ manifold showed also smaller deviations for the overall thermochemical state.
\newpage
\section{Assessment of flame-tangential diffusion effects} \label{A:tangential_diffusion}

To provide further insights into the performance of the manifolds, diffusive fluxes of selected species are evaluated with respect to the corrugated flame front of the thermo-diffusively unstable spherical expanding flame.

\begin{figure*}[ht]
\begin{center}
\includegraphics[scale=1]{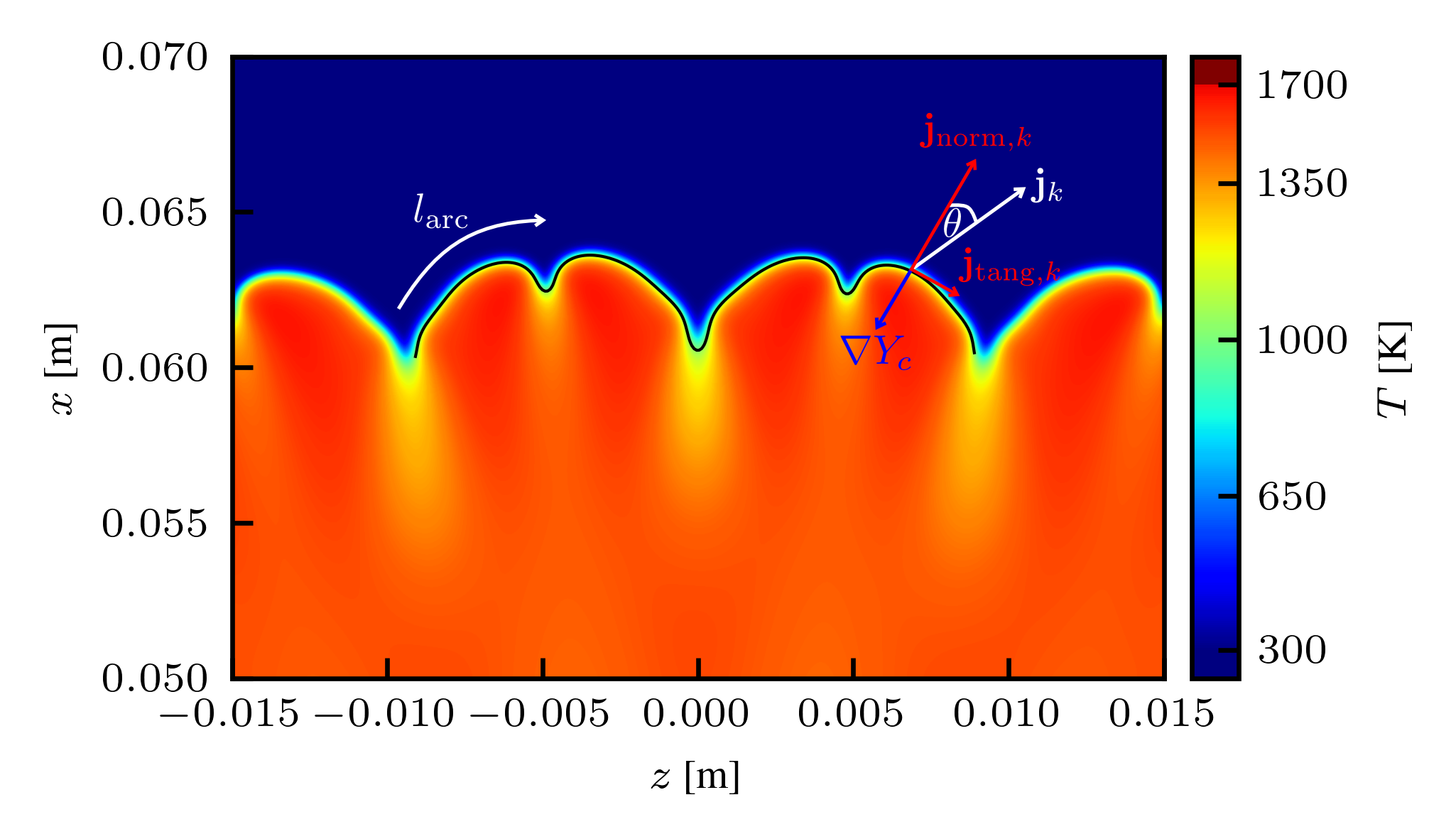}
\end{center}
\caption{Zoom into the flame front color-coded by temperature. The subset of the flame front (progress variable iso-line $Y_c=-0.11$) is shown as black line. Additionally, the arc length $l_\mathrm{arc}^{n+1}= l_\mathrm{arc}^n + \sqrt{\mathrm{d}x^2+\mathrm{d}z^2}$ is introduced as the pointwise distance of the flame front segment where $n$ represents the point index. Further, the gradient alignment between $\nabla Y_c$ and $\mathbf{j}_k$ is schematically depicted. }
\label{fig:flame_front}
\end{figure*}

For the following analysis, a segment of the flame front is considered. In \figurename\ref{fig:flame_front}, a subset of the flame front (DC reference result) is visualized and color-coded by temperature $T$. Additionally, the segment of the flame front is highlighted as black line and the arc length $l_\mathrm{arc}$ is introduced as coordinate measured along the flame front segment and subsequently used to compare tangential diffusion effects. Further, the gradient alignment between progress variable gradient $\nabla Y_c$ and diffusive flux $\mathbf{j}_k$ is depicted schematically.

The diffusive flux of species $k$ is defined as $\mathbf{j}_k=-D_{\mathrm{m},k} \nabla Y_k$, where $D_{\mathrm{m},k}$ is the mixture-averaged diffusion coefficient of species $k$. It can be further decomposed into a flame-normal and a flame-tangential component by relating it to the direction of the progress variable gradient. Therefore, the angle $\theta$ between the species gradient $\nabla Y_k$ and the progress variable gradient $\nabla Y_c$ is determined by:
\begin{equation}
    \theta=\text{arccos} \left(\frac{\nabla Y_k \cdot \nabla Y_c}{|\nabla Y_k||\nabla Y_c|}\right).
\end{equation}
This information about the gradient alignment is then used to evaluate the flame-normal diffusive flux $\mathbf{j}_{\mathrm{norm},k}=\mathbf{j}_k \, \text{cos}(\theta)$ and the flame-tangential diffusive flux $\mathbf{j}_{\mathrm{tang},k}=\mathbf{j}_k \, \text{sin}(\theta)$ of each species, respectively.

\begin{figure*}[ht]
\begin{center}
\includegraphics[scale=1]{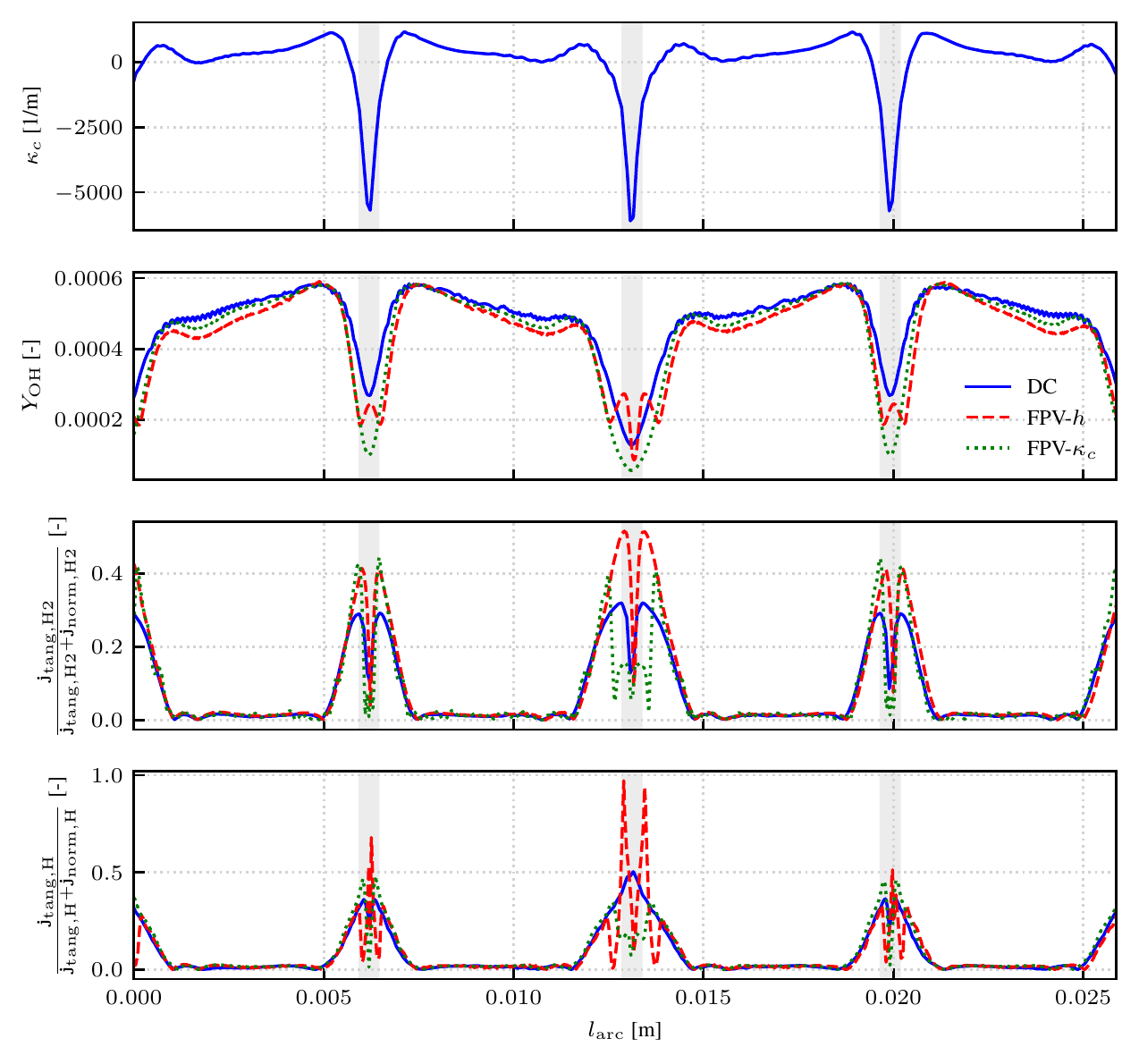}
\end{center}
    \caption{A-priori assessment of flame-tangential diffusion effects along the arc length of the flame front $l_\mathrm{arc}$ of the thermo-diffusively unstable flame. The curvature $\kappa_c$, the \ce{OH} species mass fraction $Y_\mathrm{OH}$, and the normalized tangential diffusive flux for \ce{H2} and \ce{H} are shown for the DC reference and both manifold predictions. Further, areas where the curvature exceeds the range captured by the FPV-$\kappa_c$ manifold, i.e.~regions with $\kappa_c < \,$\SI{-1500}{\per\meter}, are marked by grey shading. }
    \label{fig:tangential_diff}
\end{figure*}

In \figurename\ref{fig:tangential_diff}, the curvature $\kappa_c$, the \ce{OH} species mass fraction $Y_\mathrm{OH}$, and the contribution of flame-tangential diffusion for \ce{H2} and \ce{H} are shown along the arc length $l_\mathrm{arc}$ for the DC reference data and both FPV models. For the curvature profile distinct peaks towards negative curvatures are visible. As already discussed in the manuscript, these minima correlate with the different cells of the corrugated flame front. Further, areas with $\kappa_c < \,$\SI{-1500}{\per\meter} are highlighted by grey shading, since these highly negative curvature values exceed the curvature range captured by the FPV-$\kappa_c$ manifold. Secondly, the \ce{OH} mass fraction is compared between the three models. In general, the \ce{OH} profile shows similar characteristics as the curvature profile. 
Further, the prediction of the FPV-$\kappa_c$ manifold agrees better with the DC reference compared to the FPV-$h$ manifold, while both manifolds underpredict the \ce{OH} mass fraction in areas with very negative curvature.
Finally, the flame-tangential diffusive fluxes of \ce{H2} and \ce{H} are shown in a normalized manner. It is clearly visible that flame-tangential diffusion becomes increasingly relevant for highly negatively curved flame segments and is negligible elsewhere. The FPV models can recover flame-tangential effects to a certain extent, but for very negative curvature regions both FPV models show notable deviations.

The analysis of the flame-tangential diffusion effects further supports the findings described in Section~3.1 of the manuscript. It can be concluded, that the manifolds' ability in recovering tangential diffusion effects correlates with its overall prediction of the thermo-chemical state. A more comprehensive analysis is conceivable in form of an \textit{a-posteriori} assessment where deviations introduced by the reduced number of transported scalars could be systematically evaluated. This is subject to future work.